\documentclass[twocolumn,prb,aps,showpacs,amsmath,amssymb]{revtex4-1}

\usepackage{graphicx}
\usepackage{dcolumn}
\usepackage{bm}

\renewcommand{\vec}[1]{{\bf#1}}

\begin{document}

\title{Cherenkov sound on a surface of a topological insulator}

\author{Sergey Smirnov}
\affiliation{Institut f\"ur Theoretische Physik, Universit\"at Regensburg,
  D-93040 Regensburg, Germany}

\date{\today}

\begin{abstract}
Topological insulators are currently of considerable interest due to peculiar
electronic properties originating from helical states on their surfaces. Here
we demonstrate that the sound excited by helical particles on surfaces of
topological insulators has several exotic properties fundamentally different
from sound propagating in non-helical or even isotropic helical
systems. Specifically, the sound may have strictly forward propagation absent
for isotropic helical states. Its dependence on the anisotropy of the
realistic surface-states is of distinguished behavior which may be used as an
alternative experimental tool to measure the anisotropy strength. Fascinating
from the fundamental point of view backward, or anomalous, Cherenkov sound  is
excited above the critical angle $\pi/2$ when the anisotropy exceeds a
critical value. Strikingly, at strong anisotropy the sound localizes into a
few forward and backward beams propagating along specific directions.
\end{abstract}


\maketitle

\section{Introduction}\label{intro}
A topological insulator (TI) \cite{Hasan_2010,Qi_2011} is a system supporting
helical states \cite{Wu_2006,Xu_2006} at its edges. These states,
characterized by strong coupling between their spin degree of freedom and
direction of propagation, appear as Kramers pairs and have zero gap as a
consequence of the $T$-invariance. At the same time the bulk states have a
finite gap. Therefore, these systems represent a phase of matter with
coexisting metallic edge and insulating bulk. Importantly, the helical states
are necessarily edge states (one- or two-dimensional (2D)) of a bulk system
(two- or three-dimensional (3D)) and do not exist in truly one-dimensional or
2D systems since the $T$-invariance requires for fermions an even
number of Dirac points.

One-dimensional helical states have been experimentally implemented in
semiconductor quantum wells \cite{Koenig_2007} where the quantum spin Hall
effect has been observed in the regime of the inverted band structure
supporting dissipationless edge currents \cite{Roth_2009}.

Of particular interest for the present study are 3D TIs
supporting 2D helical states \cite{Zhang_2009}. These
surface states have been experimentally observed, {\it e.g.}, in Bi$_2$Te$_3$
\cite{Chen_2009}, where a single nondegenerate Dirac cone is located at the
$\Gamma$ point of the surface Brillouin zone. The isotropic conic dependence
of the electron energy on the momentum inherent to low-energy states breaks at
higher energies. Here cubic in momentum terms reduce the continuous rotational
symmetry down to the discrete threefold rotational symmetry. As a result, the
shape of constant energy contours becomes hexagonal \cite{Fu_2009,Liu_2010} as
observed in experiments \cite{Chen_2009}. This anisotropic energy-momentum
dependence may lead to fundamentally new behavior of physical observables. In
particular, in Ref. \onlinecite{Wang_2011} it has been predicted that the
dielectric function obtained within the random phase approximation may become
anisotropic in the momentum space.

Here we address an alternative issue related to the existence of the helical
states and explore their impact on other degrees of freedom, namely lattice
vibrations, with a special focus on the Cherenkov sound (CS) excited by
helical particles on a surface of a 3D TI.

The Cherenkov effect \cite{Cherenkov,Tamm} is a fundamental physical
phenomenon having both optic \cite{Landau_VIII} and acoustic \cite{LS}
manifestations. In particular, in the acoustic Cherenkov effect a medium emits
a forward sound, distributed within the Cherenkov cone, under the impact of an
electron whose velocity is larger than the sound velocity of this medium.
This situation may change when there appears a strong coupling
between the orbital and spin electronic degrees of freedom. It has
been shown in Ref. \onlinecite{Smirnov_2011} that in a 2D system
with the Rashba \cite{Bychkov} spin-orbit interaction electrons can excite
anomalous CS, which propagates outside the Cherenkov cone in forward and
backward directions. Here anomalous CS appears in a homogeneous system due to
interchiral transitions specific to spin-orbit coupled systems.

This outstanding property of the CS in systems with strong spin-orbit coupling
provides a platform for new applications in acoustic amplification based
currently mainly on the normal CS. These more conventional applications
include, {\it e.g.}, acoustic amplification in confined systems such as
Si/SiGe/Si heterostructures \cite{Komirenko_2000} or the Cherenkov emission in
polar bulk semiconductors \cite{Zhao_2009} such as GaAs.

It is important to mention that in optics an anomalous Cherenkov effect may
appear in the absence of spin-orbit coupling but due to strong inhomogeneity
of systems as has been achieved, {\it e.g.}, in photonic crystals
\cite{Joannopoulos}.

Turning to the CS on a surface of a 3D TI one might expect a picture similar
to the one in a 2D Rashba gas. Indeed, as a result of a strong spin-orbit
coupling, anomalous CS must appear due to interchiral transitions and
propagate forward and backward. However, this scenario cannot be
realized: since the velocity $v$ on the Dirac cone is much larger than the
sound velocity $c$, $v\gg c$, interchiral transitions do not contribute and
the CS is of pure intrachiral nature.

As has been demonstrated in Ref. \onlinecite{Smirnov_2011}, in this case there
can be excited only normal CS with the standard features: 1) the
sound is located within the Cherenkov cone whose angle $\phi_c$ is (because of
$v\gg c$) close to $\pi/2$, $\phi_c\approx \pi/2$; 2) its strictly forward
propagation is forbidden.

Here we demonstrate that the discrete threefold rotational symmetry of the
system drastically changes this standard picture and leads to fundamentally
new features of the CS propagating on a surface of a TI. Among
these features are 1) strictly forward propagation and its
remarkable and valuable for experiments dependence on the anisotropy of the
helical states; 2) anomalous propagation outside the Cherenkov cone,
{\it i.e.}, for angles $\phi>\phi_c$; 3) localization into a finite number of
normal and anomalous beams in the regime of strong anisotropy.

These remarkable features of the CS distinguish TIs from other systems such
as, {\it e.g.}, graphene. Indeed, the strictly forward sound is obviously
absent in graphene because in this system the spinor components have the same
absolute value while on a surface of a TI they have different absolute values
(see the next section). In other words, what physically distinguishes the
CS on a surface of a TI is the finite out-of-plane spin polarization of this
surface. A recent investigation of the CS in graphene and its application to
hypersonic devices may be found for example in Ref. \onlinecite{Zhao_2013}.

The paper is organized as follows. In Section \ref{phys_mod} we present a
physical model able to capture the CS propagation on a surface of a 3D
TI. Next, in Section \ref{deriv_si}, we solve this physical model and derive
the CS intensity. Its behavior is explored in Section \ref{phi_0_0} for the
case when the helical particle exciting the CS is oriented along the
$x$-axis. Section \ref{phi_0_n0} generalizes the results of Section
\ref{phi_0_0} and shows results for different orientations of the helical
particle exciting the CS. The experimental relevance of the results is
discussed in Section \ref{concl}.

\section{Physical model}\label{phys_mod}
As an application of our theory to a real physical setup, we will have in mind
helical particles on a surface of Bi$_2$Te$_3$. Additionally, we will neglect
possible sources of the particle-hole asymmetry. In this case the Hamiltonian
of helical particles has the following form \cite{Fu_2009}:
\begin{equation}
\hat{H}=v(\hat{p}_x\hat{\sigma}_y-\hat{p}_y\hat{\sigma}_x)+
\frac{\lambda}{2}(\hat{p}_+^3+\hat{p}_-^3)\hat{\sigma}_z,
\label{Ham_TI}
\end{equation}
where $\hat{p}_i$ and $\hat{\sigma}_i$ ($i=x,y$) are the momentum and
spin-$1/2$ Pauli operators, respectively,
$\hat{p}_\pm\equiv\hat{p}_x\pm i\hat{p}_y$. In Eq. (\ref{Ham_TI}) the first
term describes the isotropic Dirac cone characterized by the velocity $v$
while the second term describes the reduction of the full rotational symmetry
down to the discrete threefold rotational symmetry. The strength of this
anisotropic term is characterized by the parameter $\lambda$. For Bi$_2$Te$_3$
the values of $v$ and $\lambda$ are given in Ref. \onlinecite{Fu_2009},
$v=3.87\times 10^5\,\,\text{m/s}$, $\hbar^3\lambda=250.0\,\, \text{eV}\cdot
\text{\AA}^3$.

The Hamiltonian in Eq. (\ref{Ham_TI}) is easily diagonalized \cite{Wang_2011}
and the resulting single-particle eigenenergies and eigenstates are
\begin{equation}
\epsilon_{\vec{p}\mu}=\mu\sqrt{v^2|\vec{p}|^2+\lambda^2|\vec{p}|^6\cos^2(3\Theta_\vec{p})},
\label{Sp_ee}
\end{equation}
\begin{equation}
\varphi_{\vec{p}\mu}=
\begin{pmatrix}
\cos(\alpha_{\vec{p}\mu})\\\mu i e^{i\Theta_\vec{p}}\sin(\gamma_{\vec{p}\mu})
\end{pmatrix},
\label{es_hp}
\end{equation}
where $\mu=\pm$ and $\Theta_\vec{p}$ is the angle between the momentum
$\vec{p}$ and the $x$-axis, $\alpha_{\vec{p}\mu}\equiv(1-\mu)\pi/4-\beta_\vec{p}$,
$\gamma_{\vec{p}\mu}\equiv(1-\mu)\pi/4+\beta_\vec{p}$,
$\sin(\beta_\vec{p})=\sqrt{r(\vec{p})/[1+r(\vec{p})]}$,
and $\cos(\beta_\vec{p})=1/\sqrt{1+r(\vec{p})}$
\begin{equation}
r(\vec{p})=\frac{\sqrt{v^2p^2+\lambda^2p^6\cos^2(3\Theta_\vec{p})}-\lambda
  p^3\cos(3\Theta_\vec{p})}
{\sqrt{v^2p^2+\lambda^2p^6\cos^2(3\Theta_\vec{p})}+
  \lambda p^3\cos(3\Theta_\vec{p})}.
\label{r_func}
\end{equation}

The second quantized phonon Hamiltonian \cite{Landau_V} is
$\hat{H}_\text{ph}=\sum_{\vec{k}}\hbar\omega(\vec{k})(b^\dagger_\vec{k}b_\vec{k}+1/2)$,
where $b^\dagger_\vec{k}$, $b_\vec{k}$ are the phonon creation and
annihilation operators, respectively. We consider acoustic phonons and assume
the following phonon spectrum $\hbar\omega(\vec{k})=c|\vec{k}|$ where $c$ is
the sound velocity. In principle in Bi$_2$Te$_3$ there are longitudinal and
transverse acoustic phonons with the corresponding sound velocities,
$c_l=2.84\times 10^3\,\text{m/s}$, $c_t=1.59\times 10^3\,\text{m/s}$. However,
for simplicity, we will assume the isotropic Debye model with $c=c_l$.

The helical electrons on a surface of a TI can excite
sound. This happens via electron-phonon interaction. Due to this interaction
the medium can emit phonons at any temperature. To explore the basic features
of the CS we use the following Hamiltonian of the electron-phonon
interaction \cite{AGD},
\begin{equation}
\begin{split}
&\hat{H}_\text{el-ph}=g\sum_\sigma\int d\vec{r}\hat{\psi}^\dagger_\sigma(\vec{r})\hat{\psi}_\sigma(\vec{r})\hat{\varphi}(\vec{r}),\\
&\hat{\varphi}(\vec{r})=i\sum_{\vec{k}}\sqrt{\frac{\hbar\omega(\vec{k})}{2V}}\biggl[\exp\biggl(i\frac{\vec{k}\vec{r}}{\hbar}\biggl)b_\vec{k}-h.c.],
\end{split}
\label{Ham_eph}
\end{equation}
where $g$ is the strength of the electron-phonon interaction, $V$ is the
volume and $\hat{\psi}^\dagger_\sigma(\vec{r})$, $\hat{\psi}_\sigma(\vec{r})$
are, respectively, the helical particle creation and annihilation field operators.

\section{Derivation of the sound intensity}\label{deriv_si}
The specific nature of the CS on a surface of a 3D TI is rooted in the
properties of the eigenenergies (\ref{Sp_ee}) and eigenstates (\ref{es_hp}) of
the Hamiltonian (\ref{Ham_TI}) describing the helical particles.

\begin{figure}
\includegraphics[width=8.0 cm]{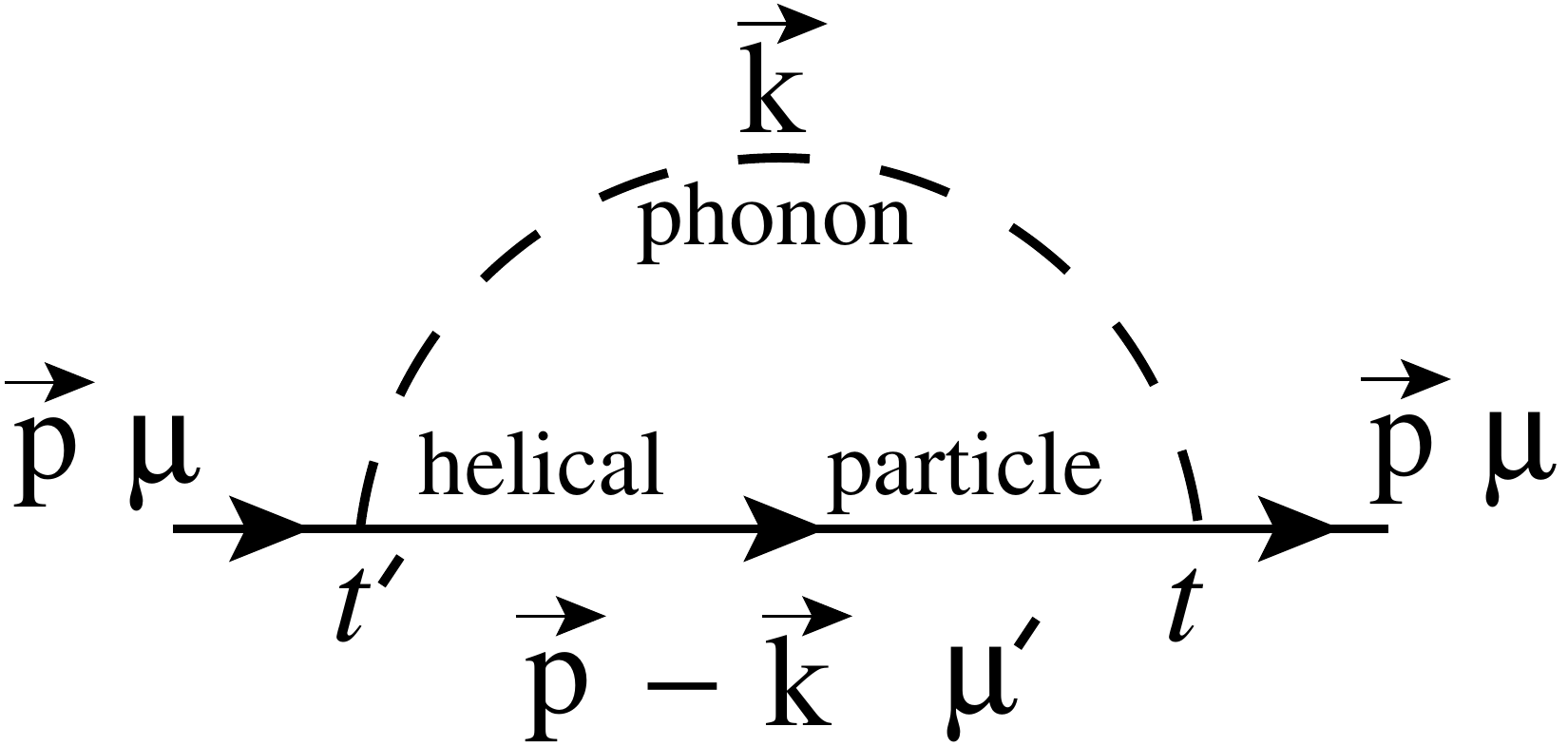}
\caption{\label{figure_1} Feynman diagram for the second-order
  contribution to the helical particle self-energy due to its interaction with
  phonons.}
\end{figure}

Using $\epsilon_{\vec{p}\mu}$ and $\varphi_{\vec{p}\mu}$ as well as the rules
for the analytic reading \cite{AGD} of Feynman diagrams, one may write down
the analytic expression corresponding to the second order (in the strength $g$
of the interaction between helical particles and phonons) diagram,
Fig. \ref{figure_1}, for the self-energy of a helical particle:
\begin{equation}
\begin{split}
&\Sigma_\mu(\vec{p},t-t')=\\
&=\frac{g^2}{\hbar}\sum_{\mu'}\int\frac{d\vec{k}}{(2\pi\hbar)^2}iG_{0\mu'}(\vec{p}-\vec{k},t-t')\times\\
&\times D_0(\vec{k},t-t')\Phi_{\mu\mu'}(\vec{p},\vec{k}),
\end{split}
\label{ep_se}
\end{equation}
where $G_{0\mu}(\vec{p},t-t')$ is the free propagator for a helical particle
with momentum $\vec{p}$ and chirality $\mu$ and $D_0(\vec{k},t-t')$ is the
free phonon propagator for a phonon with momentum $\vec{k}$. In the frequency
domain these propagators have the form:
\begin{equation}
\begin{split}
&G_{0\mu}(\vec{p},\omega)=\frac{\hbar}
{\hbar\omega-\epsilon_{\vec{p}\mu}+i\cdot 0}\\
&D_0(\vec{k},\omega)=\frac{\omega^2(\vec{k})}
{\omega^2-\omega^2(\vec{k})+i\cdot 0}.
\end{split}
\label{ep_fp}
\end{equation}
In Eq. (\ref{ep_se}) the quantity $\Phi_{\mu\mu'}(\vec{p},\vec{k})$ has the form:
\begin{equation}
\begin{split}
&\Phi_{\mu\mu'}(\vec{p},\vec{k})=\cos^2(\alpha_{\vec{p}\mu})\cos^2(\alpha_{\vec{p}-\vec{k}\mu'})+\\
&+\sin^2(\gamma_{\vec{p}\mu})\sin^2(\gamma_{\vec{p}-\vec{k}\mu'})+\\
&+2\mu\mu'\cos(\alpha_{\vec{p}\mu})\cos(\alpha_{\vec{p}-\vec{k}\mu'})\sin(\gamma_{\vec{p}\mu})\sin(\gamma_{\vec{p}-\vec{k}\mu'})\times\\
&\times\cos(\Theta_{\vec{p}-\vec{k}}-\Theta_\vec{p}).
\end{split}
\label{Phi}
\end{equation}

To get the sound intensity one has to transform Eq. (\ref{ep_se}) into the
frequency domain, {\it i.e.}, to obtain $\Sigma_\mu(\vec{p},\omega)$ and then
find its imaginary part on the mass surface, {\it i.e.},
$\text{Im}\,\Sigma_\mu(\vec{p},\omega)$ at
$\omega=\epsilon_{\vec{p}\mu}/\hbar$.

As mentioned in the Introduction, the interchiral transitions do not
contribute to the Cherenkov effect and thus it suffices to study the sound
excited by helical particles with only one chirality. Choosing $\mu=+$
(conduction band), we obtain
\begin{equation}
\begin{split}
&\text{Im}\,\Sigma_{+}(\vec{p},\omega=\epsilon_{\vec{p}+}/\hbar)=-\frac{g^2c}{8\pi\hbar^3}\times\\
&\times\int_0^{k_\text{D}} \!\!\!dk\int_{-\pi}^{\pi} \!\!\!d\phi\,
  k^2\Phi_{++}(\vec{p},\vec{k})\delta(\epsilon_{\vec{p}+}-\epsilon_{\vec{p}-\vec{k}+}-ck),
\end{split}
\label{im_se}
\end{equation}
where $k_\text{D}$ is the Debye momentum.

Denoting the angle between the $x$-axis and the momentum of the helical
particle, exciting CS, through $\phi_0$ ({\it i.e.}, $\Theta_\vec{p}=\phi_0$)
and employing the formula
\begin{equation}
\delta[h(x)]=\sum_i\frac{1}{|h'(x_i)|}\delta(x-x_i),
\label{df_r}
\end{equation}
where $h'(x)\equiv d[h(x)]/dx$ and $h(x_i)=0$, we finally obtain
\begin{equation}
\text{Im}\,\Sigma_{+}(\vec{p},\omega=\epsilon_{\vec{p}+}/\hbar)=-\frac{g^2p^2}{8\pi\hbar^3}\int_{-\pi}^{\pi}d\phi\,W(\phi).
\label{im_se_final}
\end{equation}
Defining $x\equiv k/p$  and taking into account that for a fixed momentum
$\vec{p}$ one has $\Phi_{++}(\vec{p},\vec{k})=\Phi_{++}(x,\phi)$, where $\phi$
is the angle between $\vec{p}$ and $\vec{k}$, the dimensionless sound
intensity $W(\phi)$ in Eq. (\ref{im_se_final}) may be written as follows:
\begin{equation}
W(\phi)=\sum_i\frac{x_i^2(\phi)\Phi_{++}[x_i(\phi),\phi]}{|\chi'[x_i(\phi),\phi]|},
\label{si}
\end{equation}
where $x_i(\phi)$ are the roots of the equation
$\Delta\varepsilon(x,\phi)\equiv\epsilon_{\vec{p}+}-\epsilon_{\vec{p}-\vec{k}+}-ck=0$
accounting for the energy and momentum conservation in the system. It can be
written as
\begin{equation}
\begin{split}
\sqrt{a^2+b^2\cos^2(3\phi_0)}&-\sqrt{a^2\zeta(x,\phi)+b^2\xi(x,\phi)}-\\
&-x=0,
\end{split}
\label{ec}
\end{equation}
where
\begin{equation}
\begin{split}
&\zeta(x,\phi)\equiv 1+x^2-2x\cos(\phi),\\
&\xi(x,\phi)\equiv[\cos(\phi_0)-x\cos(\phi+\phi_0)]^2\times\\
&\times[1-4\sin^2(\phi_0)+x^2(4\cos^2(\phi+\phi_0)-3)-\\
&-2x(4\cos(\phi_0)\cos(\phi+\phi_0)-3\cos(\phi))]^2.
\end{split}
\label{zeta_ksi_chi_f}
\end{equation}
The function $\chi'(x,\phi)$ is defined as
$\chi'(x,\phi)\equiv\partial_x\chi(x,\phi)$, where
$\chi(x,\phi)\equiv\Delta\varepsilon(x,\phi)/cp$. The dimensionless parameters
$a$ and $b$ characterize the ratio of the Dirac and sound velocities, $a\equiv
v/c$, and the strength of the energy-momentum anisotropy,
$b\equiv\lambda p^2/c$. 

Let us discuss the physical meaning of the quantities in the expression for
the sound intensity, Eq. (\ref{si}).

The roots $x_i(\phi)$ represent the phonon momenta allowed by the energy and
momentum conservation. Physically it is clear that larger values of the
allowed phonon momenta must result in larger values of the sound
intensity. This is mathematically expressed by the fact that the square of the
magnitude of the allowed phonon momenta is in the enumerator of
Eq. (\ref{si}).

However, different roots $x_i(\phi)$ have different physical
significance. Indeed, imagine that the magnitude of a given allowed phonon 
momentum $x_i(\phi)$ is infinitesimally shifted keeping the direction $\phi$
of this momentum fixed. Then $\Delta\varepsilon(x,\phi)$ will deviate from
zero indicating a violation of the energy and momentum conservation,
Eq. (\ref{ec}). For different $x_i(\phi)$ this deviation has different
rates. For a given $x_i(\phi)$ a slower deviation from the conservation laws
gives evidence for its greater physical significance and thus this phonon
momentum must bring a larger contribution to the sound intensity. Exactly this
physical aspect is mathematically controlled by $\chi'(x,\phi)$ in the
denominator of Eq. (\ref{si}). In particular, if there is another allowed
phonon momentum infinitesimally close to $x_i(\phi)$ then the energy and
momentum conservation will not be violated at all and the corresponding
contribution to the sound intensity will be very large.

To understand the physical meaning of $\Phi_{++}(x,\phi)$ in Eq. (\ref{si})
let us recall that the interaction between helical particles and phonons
is diagonal in spin. It is, therefore, useful to
consider an operator $\hat{O}$ diagonal in spin,
$\langle\vec{p}'\sigma'|\hat{O}|\vec{p}\sigma\rangle=\delta_{\sigma\sigma'}O_{\vec{p}'\vec{p}}$,
and calculate its matrix elements between the eigenstates given by
Eq. (\ref{es_hp}). One readily finds
$\langle\vec{p}'\mu'|\hat{O}|\vec{p}\mu\rangle=O_{\vec{p}'\vec{p}}w_{\vec{p}'\mu'\vec{p}\mu}$,
where
\begin{equation}
\begin{split}
&w_{\vec{p}'\mu'\vec{p}\mu}\equiv\cos(\alpha_{\vec{p}'\mu'})\cos(\alpha_{\vec{p}\mu})+\\
&+\mu'\mu e^{i(\Theta_\vec{p}-\Theta_{\vec{p}'})}\sin(\gamma_{\vec{p}'\mu'})\sin(\gamma_{\vec{p}\mu}).
\end{split}
\label{w_func}
\end{equation}
The quantity $|w_{\vec{p}'\mu'\vec{p}\mu}|^2$ determines the quantum
mechanical probability of the transition
$\varphi_{\vec{p}\mu}\rightarrow\varphi_{\vec{p}'\mu'}$ induced by the
perturbation $\hat{O}$. It is easy to see that
$\Phi_{\mu\mu'}(\vec{p},\vec{k})=|w_{\vec{p}-\vec{k}\mu'\vec{p}\mu}|^2$. Therefore,
the physical meaning of $\Phi_{++}(x,\phi)$ in Eq. (\ref{si}) is the quantum
mechanical probability for a helical particle to scatter within the conduction
band ($\mu=+$) from momentum $\vec{p}$ to momentum $\vec{p}-\vec{k}$, where
the angle between $\vec{p}$ and $\vec{k}$ is equal to $\phi$ and
$k=px$. Equivalently, this probability may be called phonon emission
probability. It is physically clear that the phonons whose emission
probability is higher will produce larger contributions to the sound
intensity. Mathematically this is expressed by the fact that
$\Phi_{++}(x,\phi)$ enters the enumerator of Eq. (\ref{si}).

\begin{figure}
\includegraphics[width=8.0 cm]{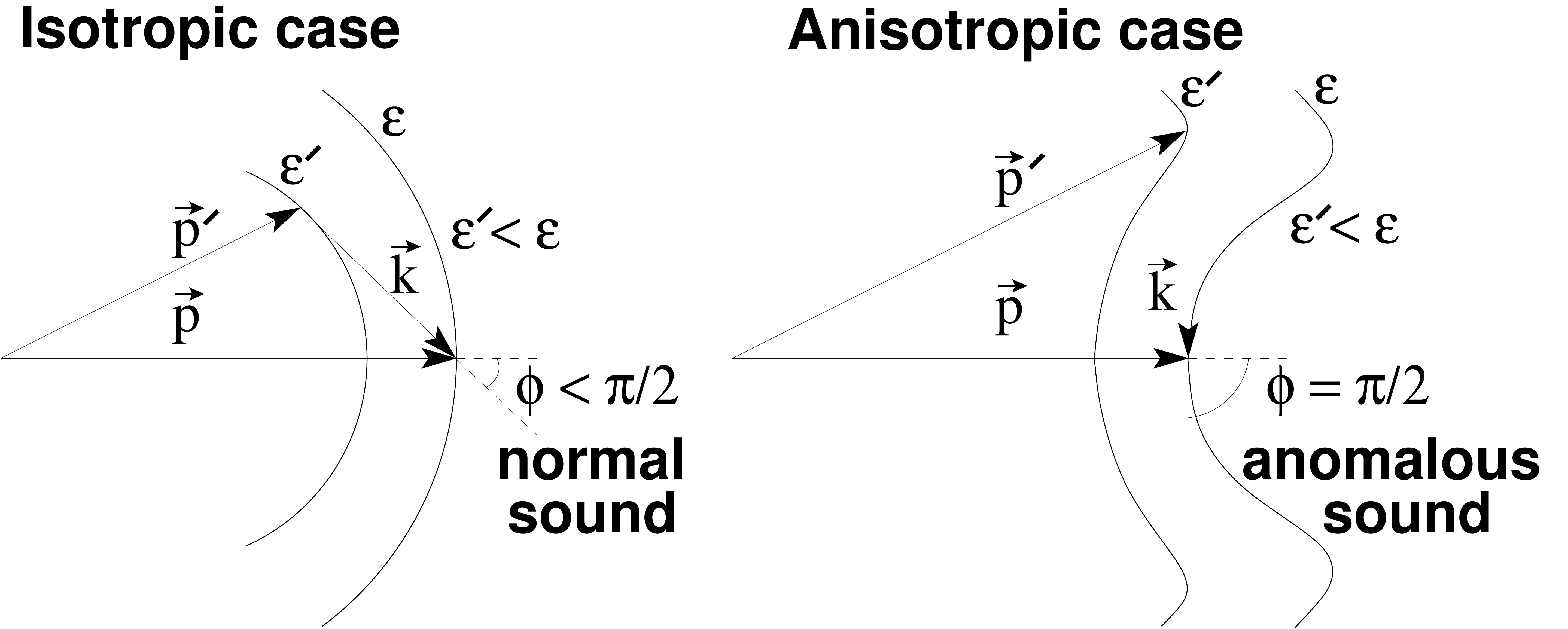}
\caption{\label{figure_2} The schematic picture of the transition
  processes allowed by the energy and momentum conservation. Here
  $\varepsilon$ and $\vec{p}$ are the energy and momentum of a helical
  particle before its scattering while $\varepsilon'$ and $\vec{p}'$ are the
  energy and momentum of this helical particle after its scattering. The
  momentum of the emitted phonon is denoted through $\vec{k}$.}
\end{figure}

Finally, Fig. \ref{figure_2} explains the physical origin of the
anomalous CS. As one can see, the reason for this sound is the anisotropy of
the constant energy surfaces. In the isotropic case an emitted phonon (with
the energy $\varepsilon-\varepsilon'$) always has its momentum $\vec{k}$ with
a forward orientation, $\phi<\pi/2$. However, when the anisotropy becomes
strong enough, it admits phonons with orientations $\phi=\pi/2$ as well as
$\phi>\pi/2$.

\section{Results for CS excited by helical particles with $\phi_0=0$}\label{phi_0_0}
All specific features of the CS on a surface of a 3D TI, mentioned in the
Introduction, may already be observed when the helical particle, exciting the
CS, is oriented along the $x$-axis, $\phi_0=0$. Therefore in this section we
consider this specific case in detail. The generalization to $\phi_0\neq 0$ is
given in the next section.
\begin{figure}
\includegraphics[width=8.0 cm]{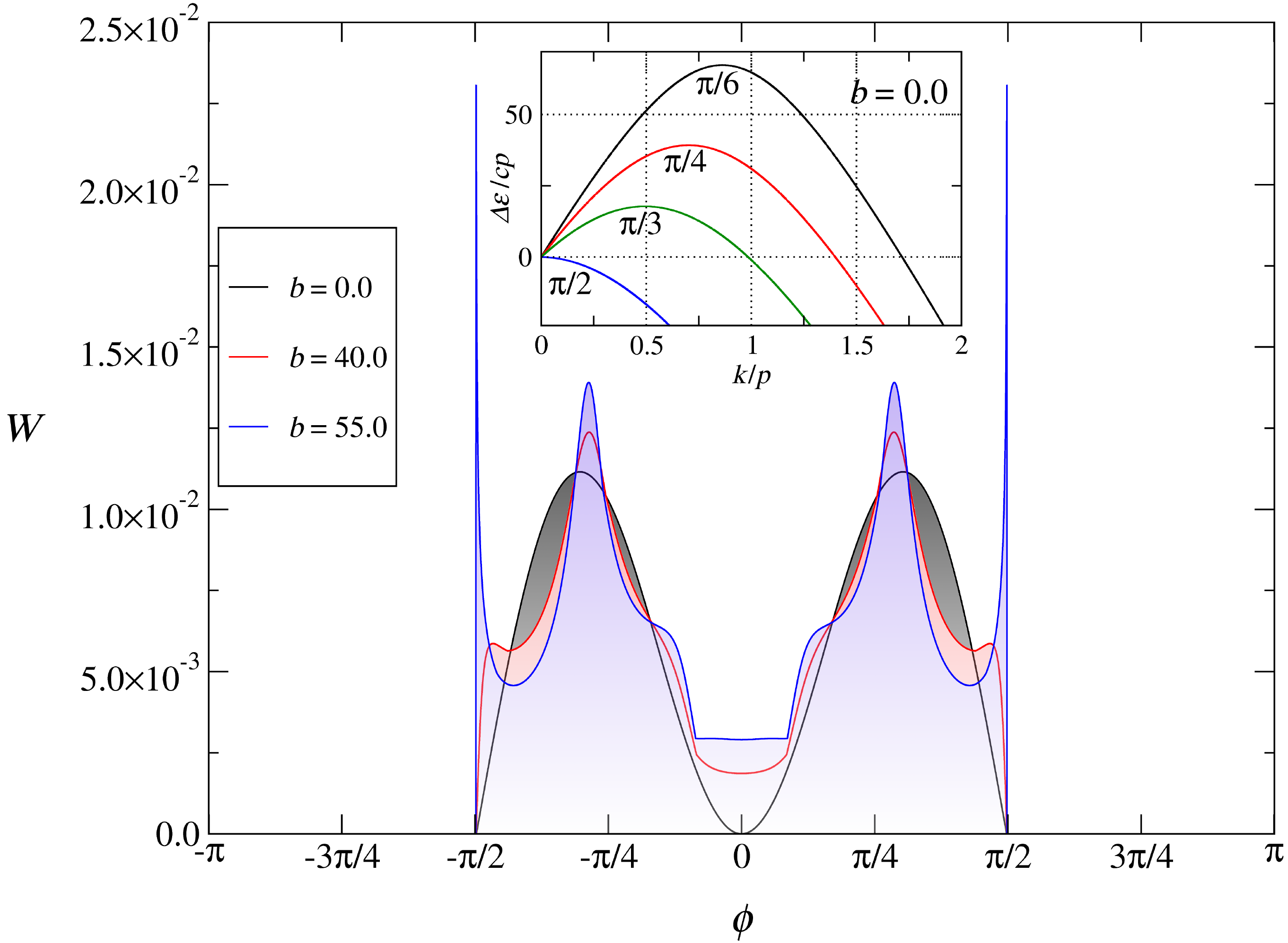}
\caption{\label{figure_3} (Color online) The CS intensity (excited by a helical
  particle moving along the $x$-axis) as a function of $\phi$ for several
  values of $b$. For Bi$_2$Te$_3$ $a=136.3$. Inset: $\Delta\varepsilon/cp$ as
  a function of $k/p$ for different angles $\phi$ and $b=0$.}
\end{figure}
\begin{figure}
\includegraphics[width=8.0 cm]{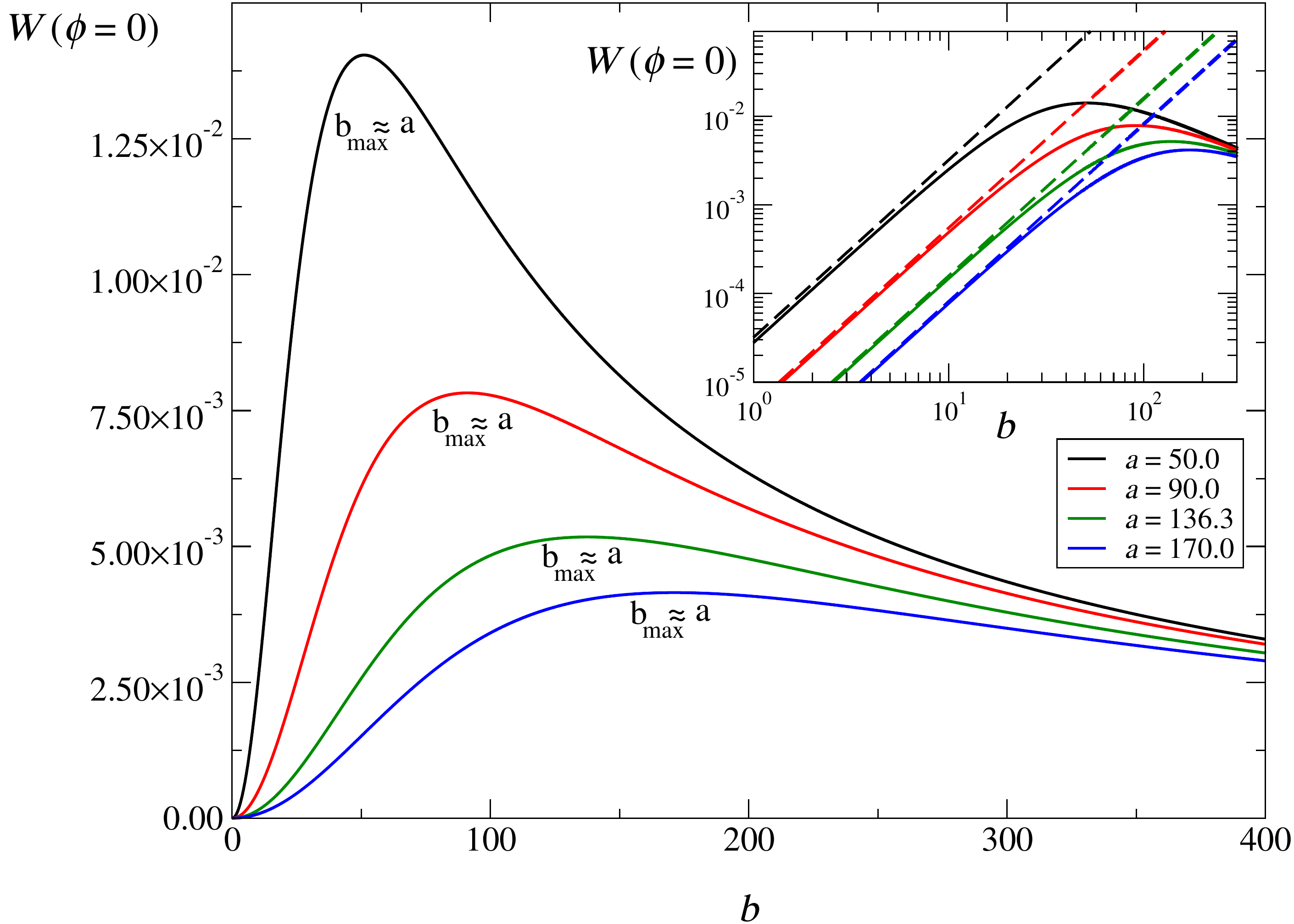}
\caption{\label{figure_4} (Color online) The intensity of the strictly forward
CS as a function of the anisotropy strength $b$ for several
values of the parameter $a$. The inset shows the expected power law behavior
(which is linear in the log-log scale) at small $b$.}
\end{figure}
\begin{figure}
\includegraphics[width=8.0 cm]{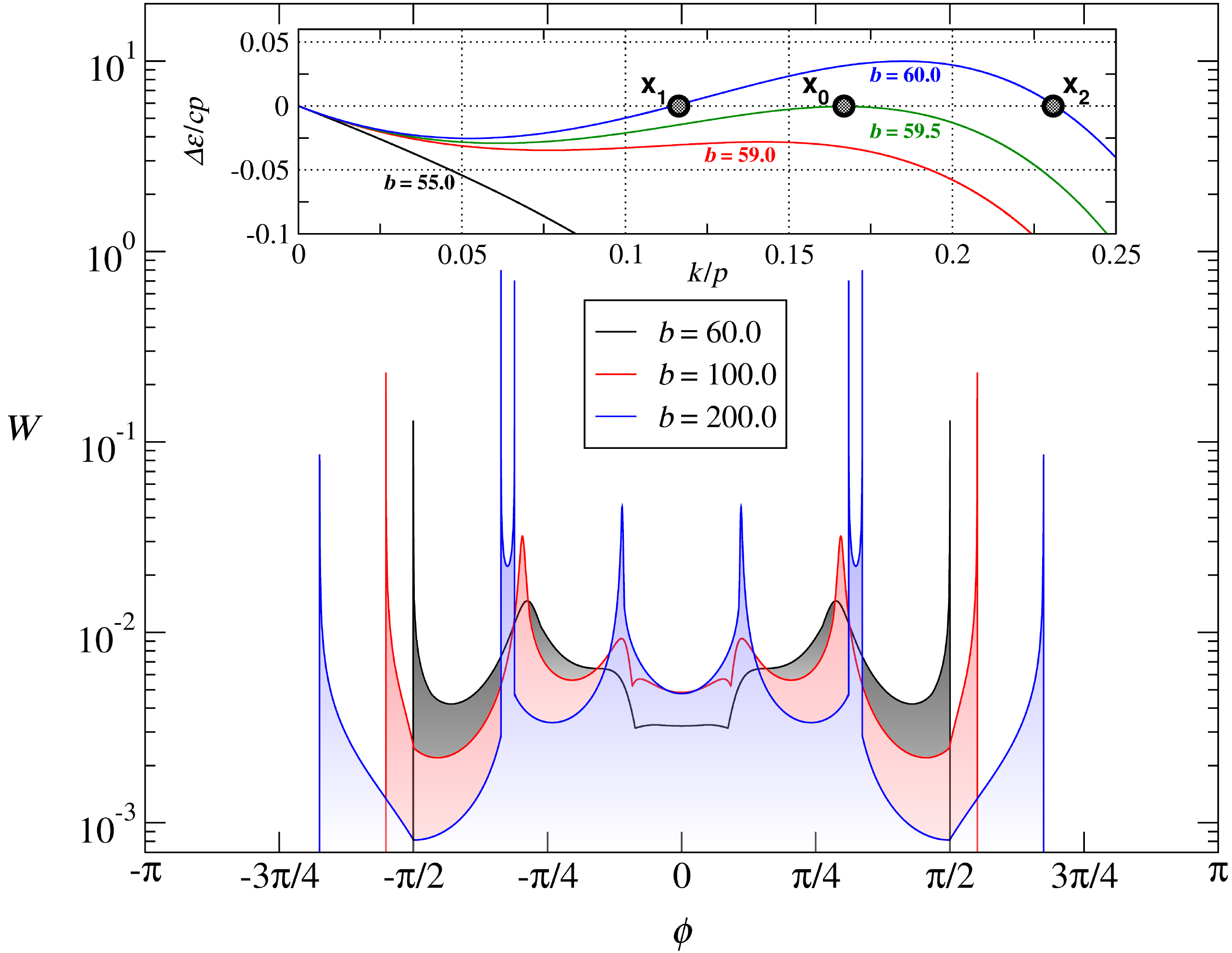}
\caption{\label{figure_5} (Color online) The angular distribution of the
  CS intensity for stronger anisotropy $b$. Here $a=136.3$. Inset:
  $\Delta\varepsilon/cp$ as a function of $k/p$ at $\phi=\phi_c$ and $b=55.0$
  (black), $b=59.0$ (red), $b=59.5$ (green) and $b=60.0$ (blue).}
\end{figure}
\begin{figure}
\includegraphics[width=8.0 cm]{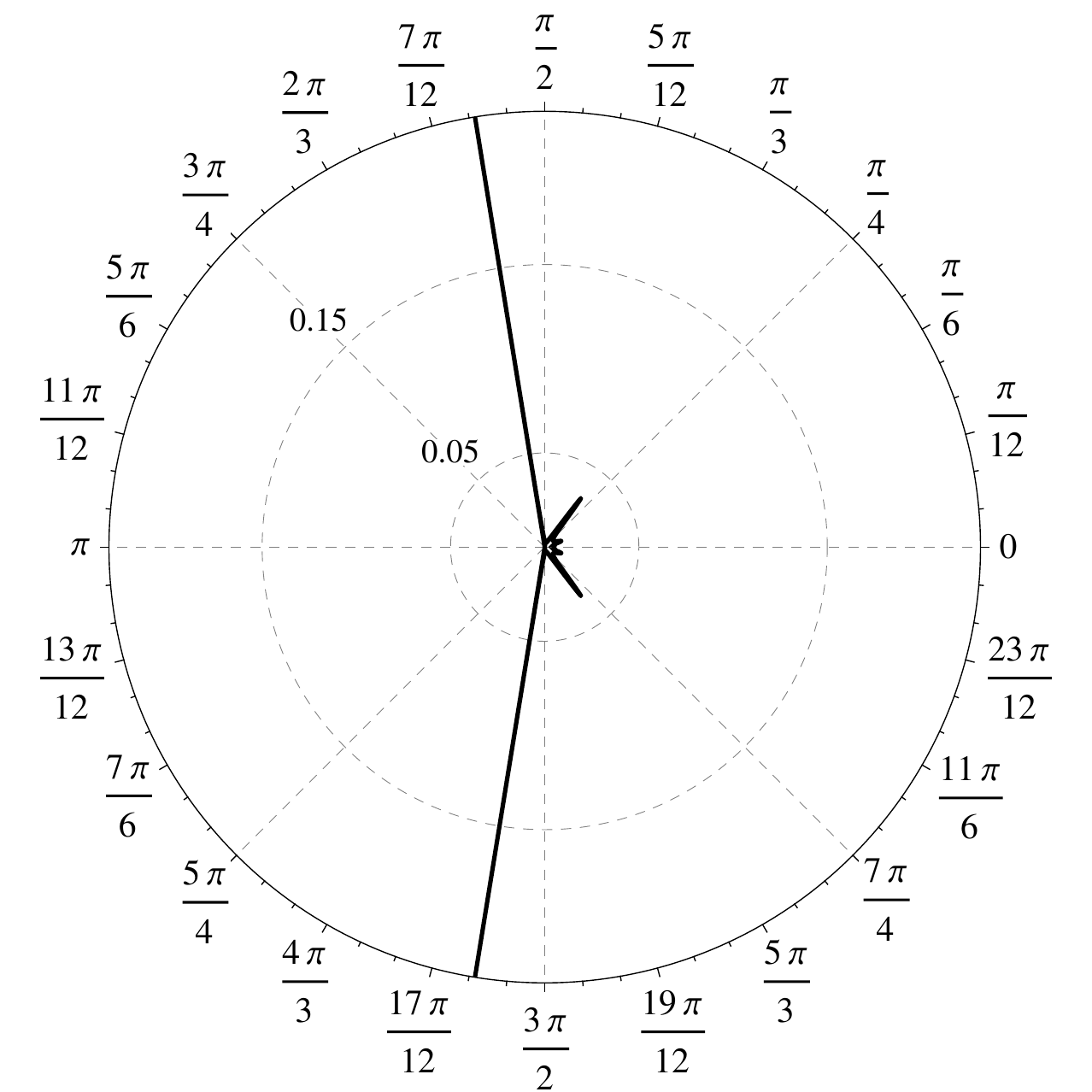}
\caption{\label{figure_6} The 2D distribution of the CS on a surface of a TI
  for $a=136.3$ and $b=100.0$.}
\end{figure}

The angular distribution of the CS intensity for not too large values of the
anisotropy parameter $b$ is shown in Fig. \ref{figure_3}. When the anisotropy
increases there appears a plateau at small angles and sharp peaks on the
surface of the Cherenkov cone $\phi=\pm\phi_c\approx \pm\pi/2$. Physically,
this behavior can be explained in terms of the quantum mechanical probability
of a phonon emission and the energy and momentum conservation (see the
previous section). When the angle increases from zero to a small finite value,
the phonon emission probability, obtained from Eq. (\ref{w_func}),
decreases. But phonon momenta with finite angles bring larger contributions to
the sound intensity. As explained in Section \ref{deriv_si}, for these momenta
the violation of the energy and momentum conservation (when infinitesimally
shifting their magnitudes and keeping their orientations unchanged) becomes
much weaker when the angle grows (this fact is mathematically controlled by
the denominator in Eq. (\ref{si})). This compensates the decrease of the
emission probability leading to a plateau. At angles close to $\phi_c$ a
finite momentum, allowed by the energy and momentum conservation, approaches
the zero momentum (also allowed). Therefore, an infinitesimal shift of its
magnitude has a little impact on the energy and momentum conservation. Thus
this phonon momentum brings a very large contribution to the sound
intensity. At the same time its magnitude approaches zero. This results in a
sharp maximum in the vicinity of $\phi_c$.
\begin{figure}
\includegraphics[width=8.0 cm]{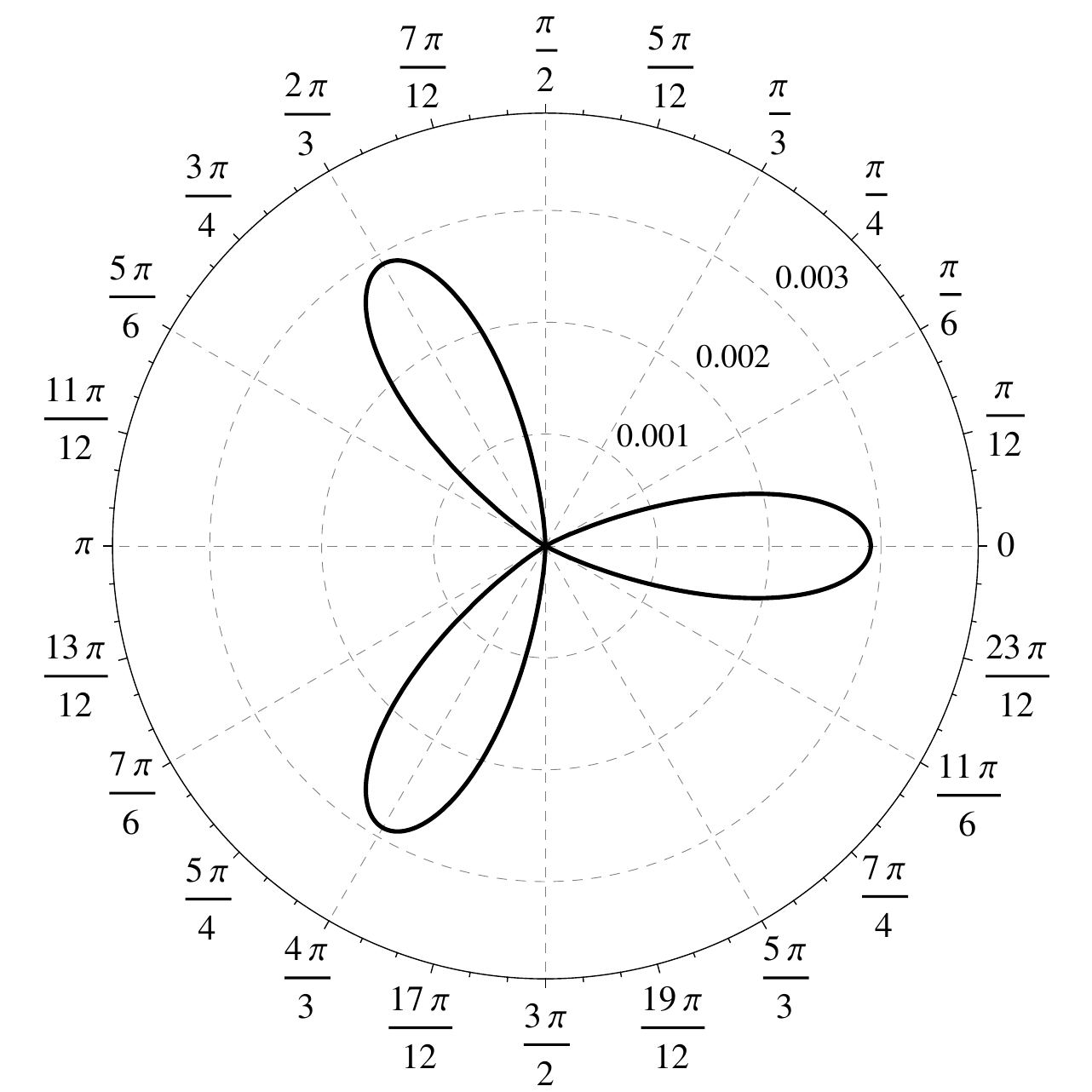}
\caption{\label{figure_7} The strictly forward CS as a function of the
  momentum orientation $\phi_0$ of the helical particle exciting the
  sound. Here $a=136.3$ and $b=55.0$.}
\end{figure}
\begin{figure}
\includegraphics[width=8.0 cm]{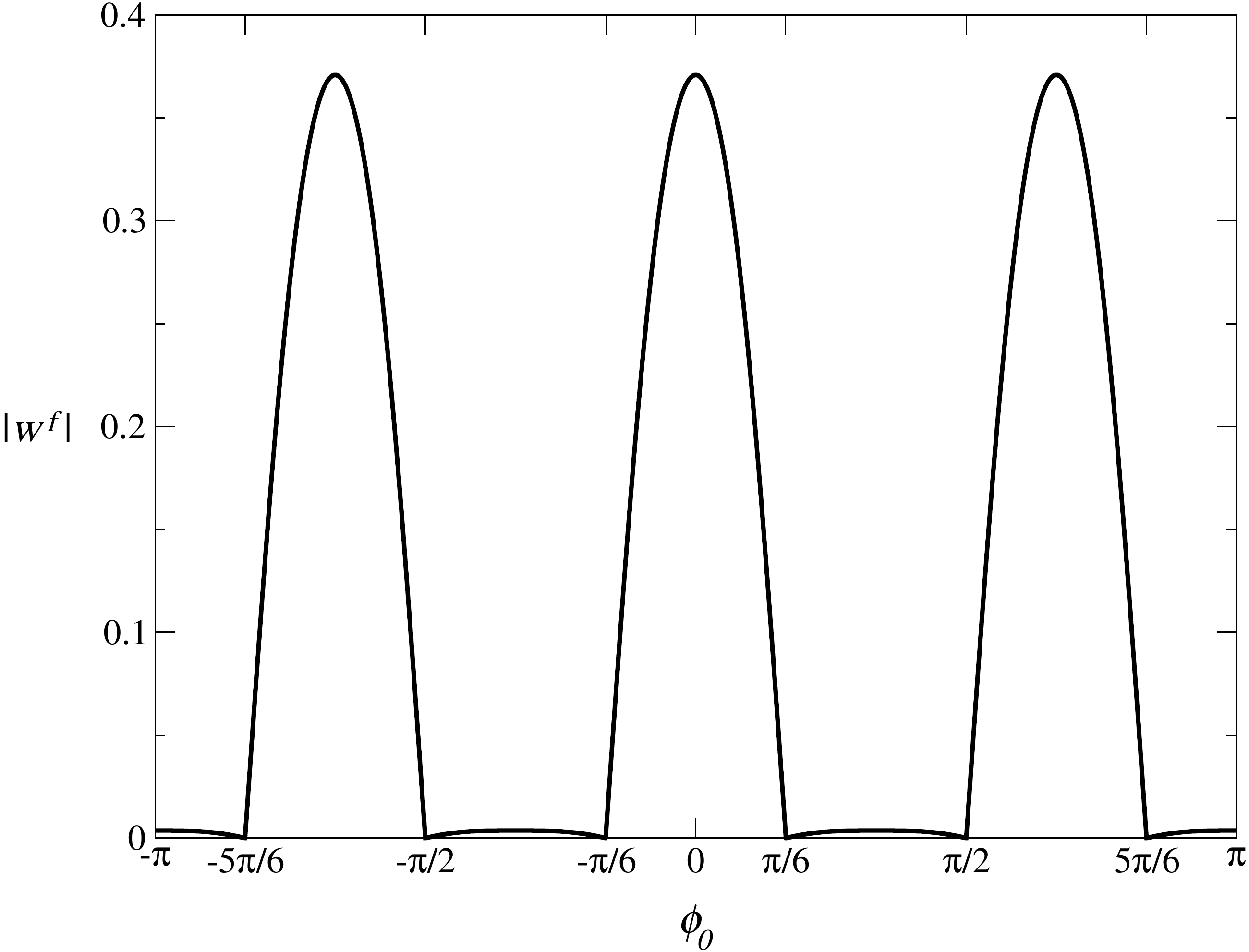}
\caption{\label{figure_8} The quantum mechanical transition probability
  for the case of the strictly forward CS as a function of the momentum
  orientation $\phi_0$ of the helical particle exciting this sound. Here
  $a=136.3$ and $b=55.0$.}
\end{figure}

The inset explains, for the case $b=0$, the mechanism of the disappearance of
the CS outside the Cherenkov cone: at angles $|\phi|<\phi_c$ the equation
\begin{figure}
\includegraphics[width=8.0 cm]{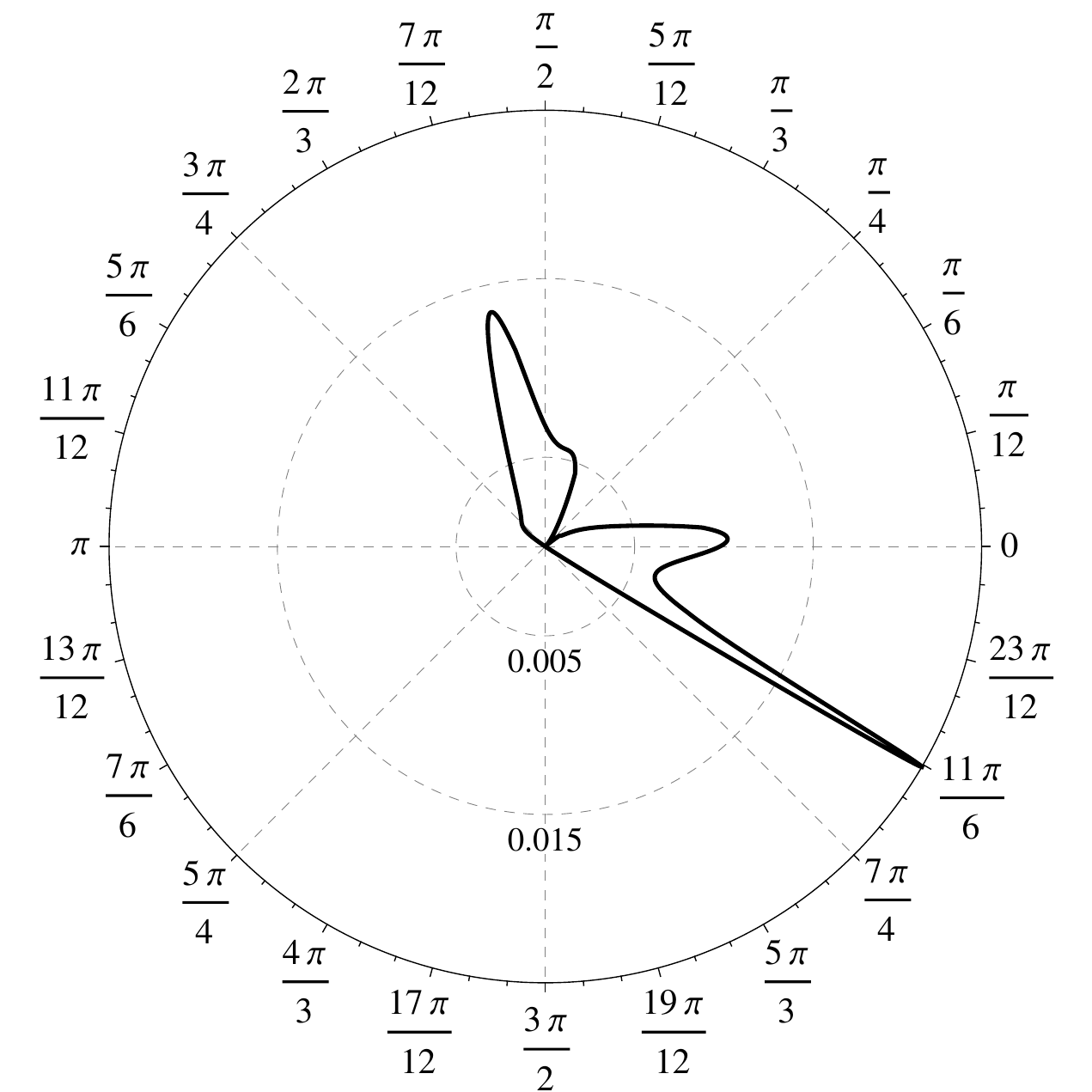}
\caption{\label{figure_9} The intensity of the CS as a function of
  $\phi+\phi_0$, where $\phi_0$ is the orientation of the helical particle
  exciting the sound, {\it i.e.}, the orientation of $\vec{p}$, while $\phi$ is
  the angle between $\vec{p}$ and the direction of the excited sound, {\it
    i.e.}, the angle between $\vec{p}$ and $\vec{k}$. Here $\phi_0=\pi/4$,
  $a=136.3$ and $b=55.0$.}
\end{figure}
\begin{figure}
\includegraphics[width=8.0 cm]{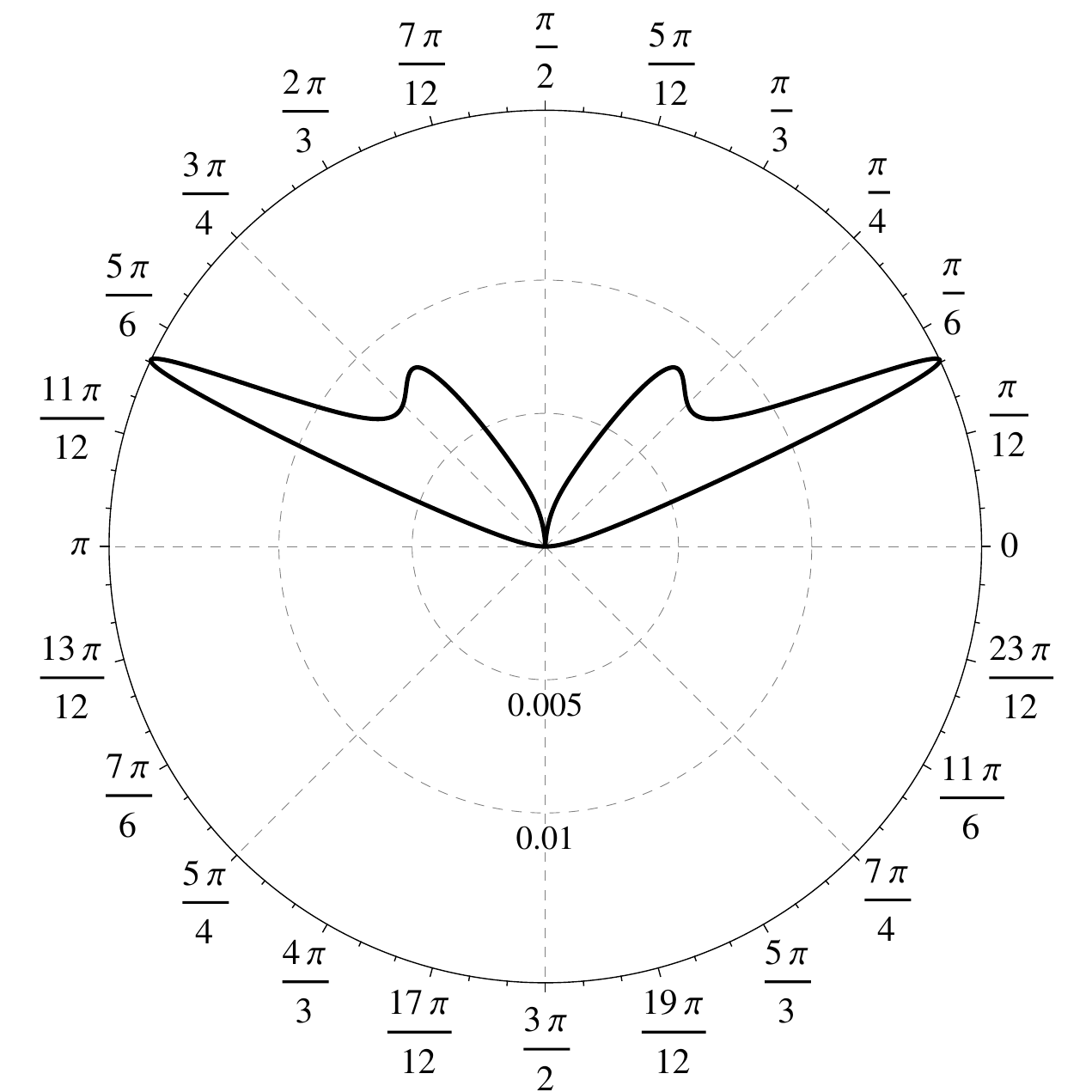}
\caption{\label{figure_10} The same as in Fig. \ref{figure_9}, but
  for $\phi_0=\pi/2$.}
\end{figure}
\begin{figure}
\includegraphics[width=8.0 cm]{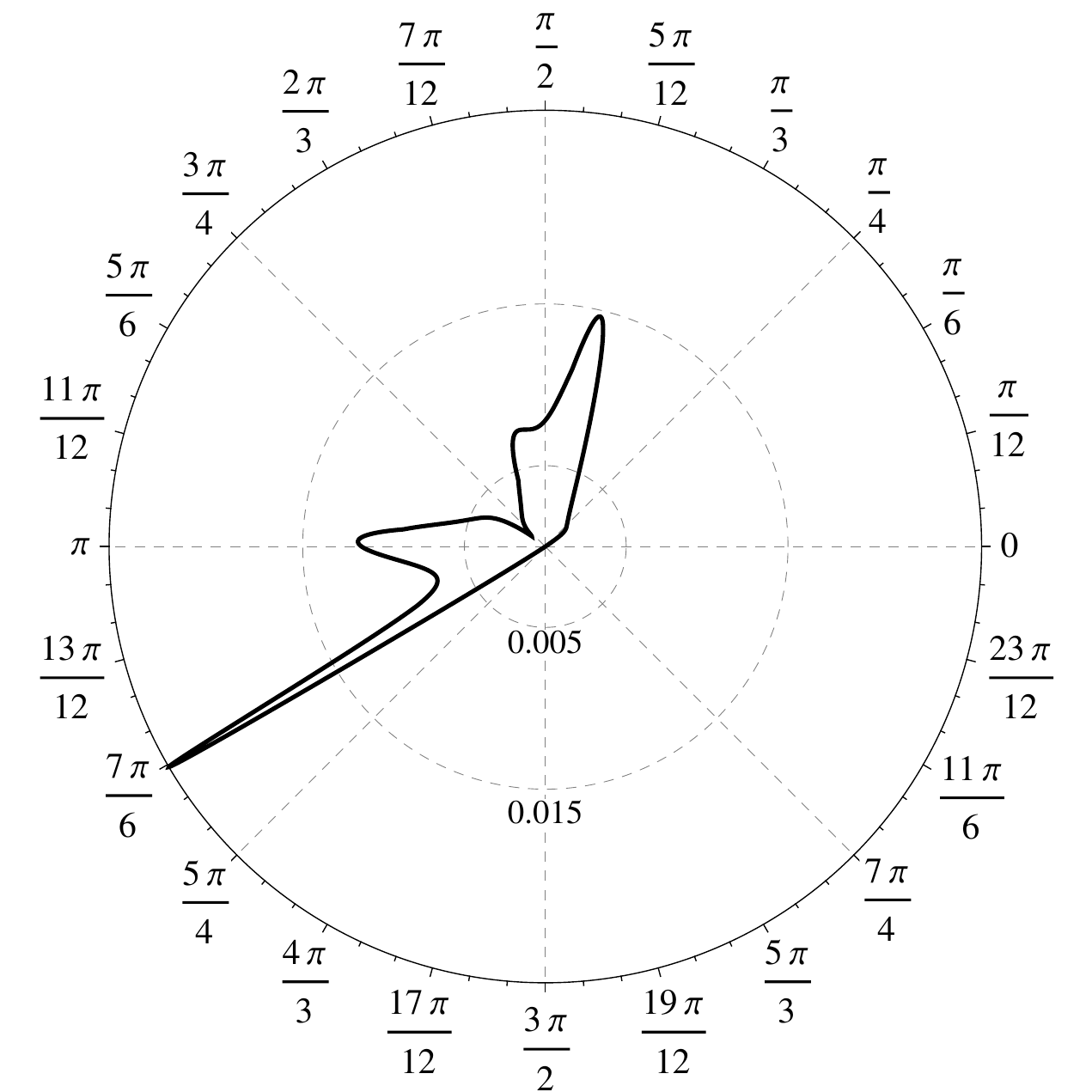}
\caption{\label{figure_11} The same as in Fig. \ref{figure_9}, but
  for $\phi_0=3\pi/4$.}
\end{figure}
\begin{figure}
\includegraphics[width=8.0 cm]{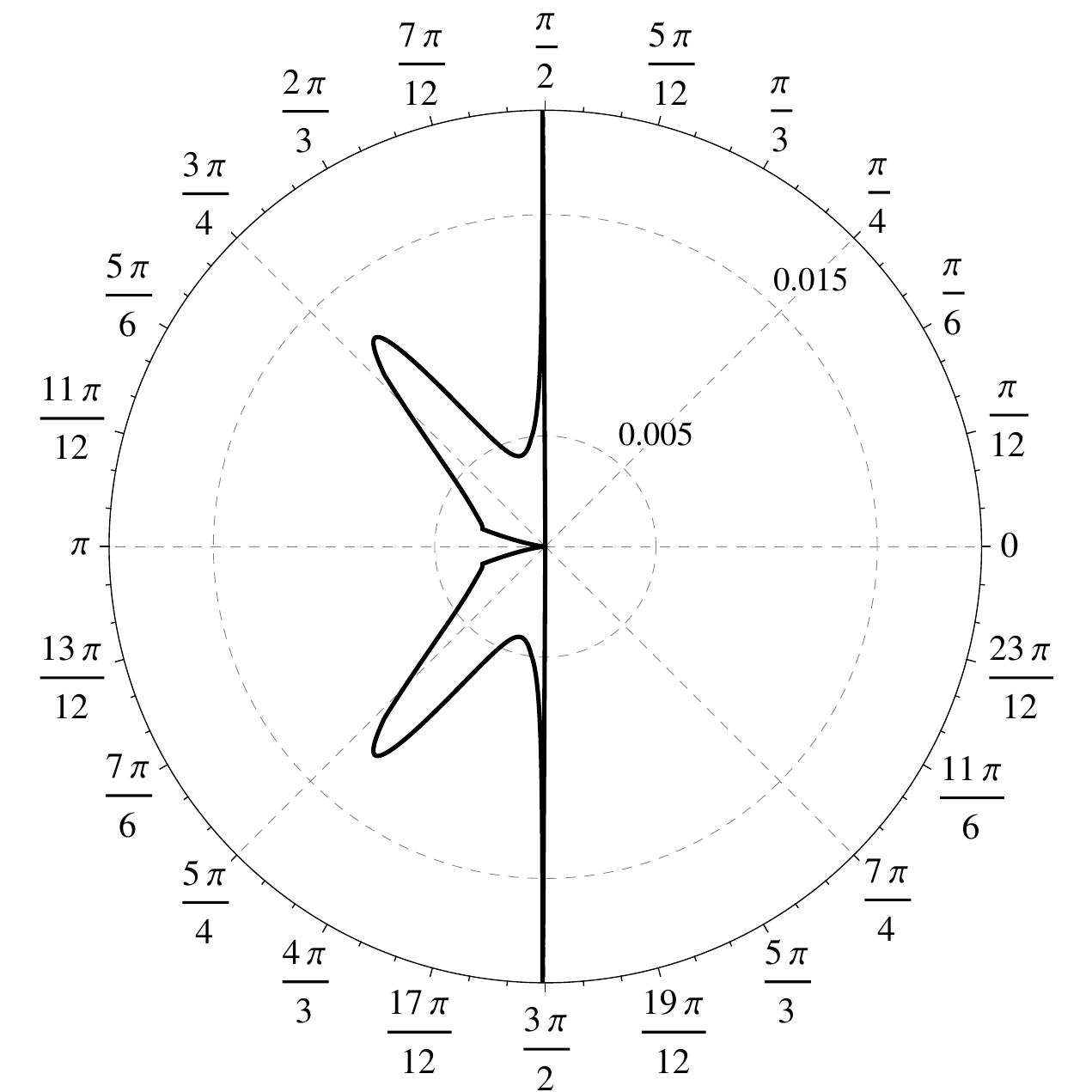}
\caption{\label{figure_12} The same as in Fig. \ref{figure_9}, but
  for $\phi_0=\pi$.}
\end{figure}
$\Delta\varepsilon/cp=0$ has one finite root $k/p$ which merges with the zero
root as soon as the angle approaches the critical value $\phi_c$. A similar
mechanism underlies the angular distribution of the sound intensity for the
case $b\neq 0$: there still exists only one finite root which merges with zero
when $\phi$ approaches $\phi_c$ but the larger $b$ is the slower this merging
becomes. Physically, the restriction of the sound to the cone can be seen from
the energy and momentum conservation. The energy of the helical particle
exciting the CS decreases. For weak anisotropy the constant energy surfaces
are almost circles. Thus the helical particle moves from a circle with a
larger radius to a circle with a smaller radius. This can only 
lead to excitation of forward sound (see Fig. \ref{figure_2}).

In Fig. \ref{figure_4} a new feature of the CS on a surface of a 3D TI,
strictly forward sound, is shown in detail. To demonstrate its properties we
plot it as a function of the anisotropy for different ratios of the Dirac and
sound velocities. All the curves have a maximum at a certain value
$b_\text{max}$. Our results clearly show that $b_\text{max}\approx a$. This is
an important issue for experiments since it implies the ratio $v=\lambda
p^2$. This ratio shows that if the momentum $p$ of the helical electron,
exciting the CS, is known, then the anisotropy parameter $\lambda$ can easily
be obtained. This is experimentally relevant because electrons exciting the
sound may be prepared with a definite momentum before they hit the
surface. Another aspect of the strictly forward sound is that at zero
anisotropy it vanishes. The inset compares this vanishing behavior with the
exact asymptotics $W(\phi=0)=4b^2/a^3=4\lambda^2p^4c/v^3$ which offers an
alternative possibility to measure the anisotropy at small momenta $p$ of
incident electrons.
\begin{figure}
\includegraphics[width=8.0 cm]{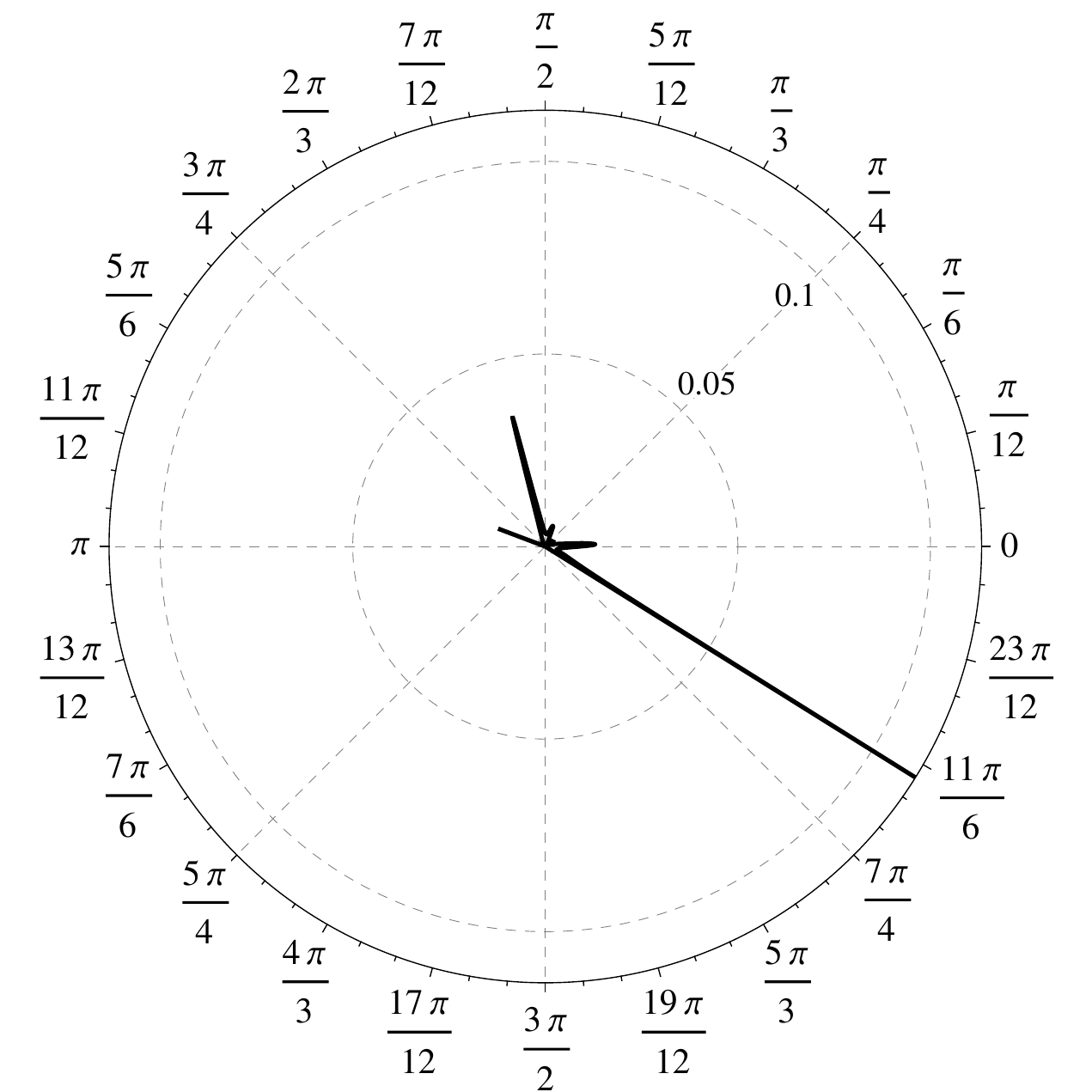}
\caption{\label{figure_13} The same as in Fig. \ref{figure_9}, but
  for $\phi_0=\pi/4$ and $b=100$.}
\end{figure}
\begin{figure}
\includegraphics[width=8.0 cm]{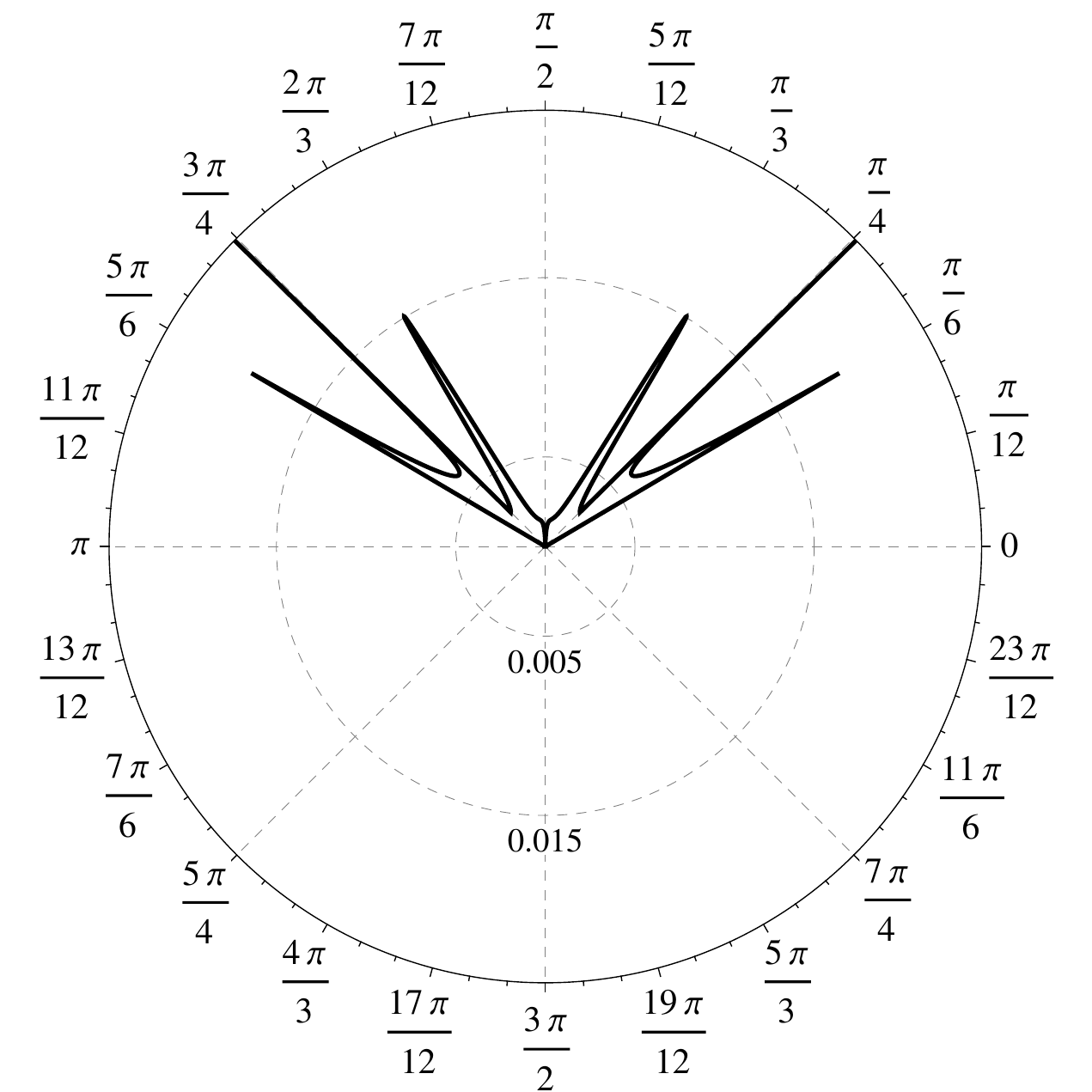}
\caption{\label{figure_14} The same as in Fig. \ref{figure_9}, but
  for $\phi_0=\pi/2$ and $b=200$.}
\end{figure}
\begin{figure}
\includegraphics[width=8.0 cm]{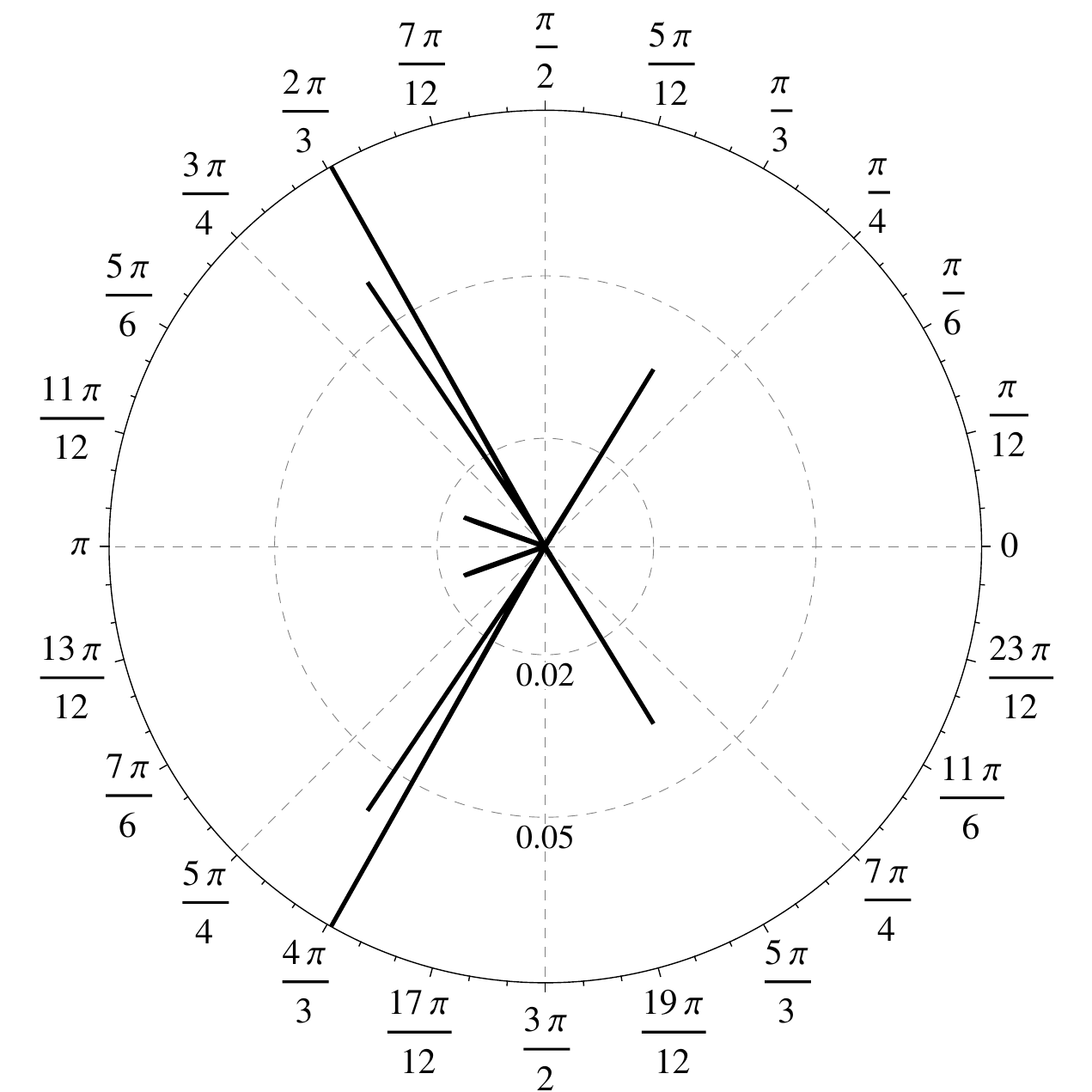}
\caption{\label{figure_15} The same as in Fig. \ref{figure_9}, but
  for $\phi_0=\pi$ and $b=200$.}
\end{figure}

The nature of the CS on a surface of a 3D TI acquires another fundamental
change when the anisotropy becomes strong as it is shown in
Fig. \ref{figure_5}. As soon as the anisotropy exceeds a critical value
$b_c=59.5$, the CS overcomes the critical angle $\phi_c\approx \pi/2$ and
starts to propagate in backward directions, 
{\it i.e.}, there appears the anomalous CS. The inset explains
the mechanism responsible for the formation of the anomalous CS. At
$\phi=\phi_c$ and $b<b_c$ the equation $\Delta\varepsilon/cp=0$ has
no finite roots $k/p$ and thus the sound intensity is zero. However, at
$b=b_c$ a single finite root appears giving a finite contribution to the sound
intensity. This leads to a jump from zero to a finite value of the sound
intensity on the surface of the Cherenkov cone $\phi=\phi_c$. Further this
root splits into two finite roots which both bring finite contributions to the
anomalous CS. The physical explanation of the anomalous CS is again given by
the energy and momentum conservation. The helical particle exciting the CS
moves from a constant energy surface with a larger energy to a constant energy
surface with a smaller energy. At the same time for strong anisotropy the
constant energy surfaces acquire a negative curvature. Exactly this negative
curvature admits the anomalous CS (see Fig. \ref{figure_2}).

The 2D distribution of the CS on a surface of a TI in the regime of strong
anisotropy is shown in Fig. \ref{figure_6} for $b=100.0$. As one can see, in
this regime the CS intensity is mainly located along specific forward and
backward directions. In other words, the CS localizes into a few normal and
anomalous beams. The physical reason for the localization of the CS at
discrete angles is that at these angles the violation of the energy and
momentum conservation (see the previous section) becomes very weak because
in the vicinities of these angles the curvature of the constant energy
surfaces becomes minimal. In the regime of strong anisotropy the angular
domains, where this curvature is minimal, become extremely narrow and, as a
result, large values of the sound intensity are located in very small areas
around discrete angles.

\section{Results for CS excited by helical particles with $\phi_0\neq 0$}\label{phi_0_n0}
Here we show some results for the CS excited by helical particles whose
momentum orientation differs from the one of the $x$-axis, that is by helical
particles with $\phi_0\neq 0$.

The results of Section \ref{deriv_si} are valid for any angle $\phi_0$ and the
sound intensity can be obtained from Eq. (\ref{si}).

In particular, one can obtain the strictly forward sound as a function of
$\phi_0$. It is shown in Fig. \ref{figure_7} using polar coordinates for
the anisotropy strength $b=55.0$ (here and below $a=136.3$ which is the value
for Bi$_2$Te$_3$). As one can see, the intensity of the strictly forward sound
has the discrete threefold rotational symmetry of the helical particle
Hamiltonian, Eq. (\ref{Ham_TI}). The specific feature of the strictly forward
sound is that it is enhanced in the sectors $(-\pi/6,\pi/6)$, $(\pi/2,5\pi/6)$
and $(-5\pi/6,-\pi/2)$ but suppressed outside them. At the angles
$\phi_0=\pm\pi/6,\pm 5\pi/6,\pm\pi/2$ the strictly forward sound vanishes
because at these values the anisotropy has no effect, as one can see from
Eqs. (\ref{Sp_ee}) and (\ref{r_func}). The enhancement and suppression of
the strictly forward sound can be explained by the behavior of the quantum
mechanical transition probability which is determined by
$w_{\vec{p}'\mu'\vec{p}\mu}$, Eq. (\ref{w_func}). In the case of the strictly
forward phonon emission it takes the form
$w^f(\vec{p},\vec{k})=\cos(\beta_{\vec{p}-\vec{k}})\cos(\beta_\vec{p})\pm\sin(\beta_{\vec{p}-\vec{k}})\sin(\beta_\vec{p})$,
where the sum is taken for $x<1$ ($k<p$) and the difference is taken for $x>1$
($k>p$). From this expression it is also easy to see that in the absence of
the anisotropy, $b=0$, $w^f(\vec{p},\vec{k})=0$. Indeed, for $b=0$ and
$a\gg 1$ (which is our case because $a=136.3$) the only finite solution
allowed by the energy and momentum conservation is $x\approx
2-2/a$. Therefore, $x>1$ and
$w^f(\vec{p},\vec{k})=\cos(\beta_{\vec{p}-\vec{k}})\cos(\beta_\vec{p})-\sin(\beta_{\vec{p}-\vec{k}})\sin(\beta_\vec{p})$. Since
for $b=0$ we have
$\cos(\beta_{\vec{p}-\vec{k}})=\cos(\beta_\vec{p})=\sin(\beta_{\vec{p}-\vec{k}})=\sin(\beta_\vec{p})=1/\sqrt{2}$,
we conclude that $w^f(\vec{p},\vec{k})=0$. Therefore, the reason for nonzero
strictly forward sound is the anisotropic nature of the helical states. At
finite anisotropy, $b=55.0$, the absolute value $|w^f(\vec{p},\vec{k})|$ is
shown in Fig. \ref{figure_8} as a function of $\phi_0$. It has large
values in the sectors $(-\pi/6,\pi/6)$, $(\pi/2,5\pi/6)$ and
$(-5\pi/6,-\pi/2)$ and it is very small outside them. This nicely explains the
specific dependence of the strictly forward sound shown in
Fig. \ref{figure_7}.

Finally, we show the angular distribution of the CS intensity on a surface of
a 3D TI for different orientations of the helical particle exciting the sound,
{\it i.e.}, for different values of $\phi_0$ as well as for different levels
of the anisotropy, {\it i.e.}, for different values of $b$.

In Fig. \ref{figure_9} the CS intensity is shown for $\phi_0=\pi/4$ and
$b=55.0$. The characteristic feature of the sound distribution in this case is
its asymmetry with respect to the orientation of the helical particle, {\it
  i.e.}, with respect to $\vec{p}$. The symmetry of the sound distribution is
restored when $\vec{p}$ approaches an orientation with respect to which the
constant energy surfaces have a certain symmetry as it happens, {\it e.g.}, in
the case $\phi_0=\pi/2$ shown in Fig. \ref{figure_10}. Increasing $\phi_0$
further again leads to a loss of the symmetry of the sound distribution, shown
in Fig. \ref{figure_11} for $\phi_0=3\pi/4$. Another recovery of the
symmetry takes place at $\phi_0=\pi$, Fig. \ref{figure_12}.

In the previous section it has been shown that at strong anisotropy a helical
particle moving along the $x$-axis, $\phi_0=0$, excites the CS propagating
mainly along specific directions, {\it i.e.}, the CS localizes into a few
forward and backward beams. Here we show that this specific feature of the
CS on a surface of a 3D TI is retained when the particle exciting the sound
moves along an arbitrary direction $\phi_0\neq 0$. Indeed,
Fig. \ref{figure_13} shows for the case $\phi_0=\pi/4$ and $b=100$ that
the sound is mainly located within four forward beams and one backward
beam. Fig. \ref{figure_14} demonstrates that in the case $\phi_0=\pi/2$,
$b=200$ there are only six forward beams, which are a bit delocalized, while
for $\phi_0=\pi$, $b=200$ Fig. \ref{figure_15} illustrates a strong
localization of the CS into eight beams, six forward and two backward ones.

\section{Conclusions}\label{concl}
In conclusion, let us estimate the relevance of our results for
experiments. In the regime of strong anisotropy, {\it e.g.}, at $b=200.0$, one
gets $p\approx 1.28\cdot 10^{-25}$ kg$\cdot$m/s. In terms of the corresponding
wave vector, $p=\hbar q$, one has $q\approx 0.12\AA^{-1}$ which is well within
the modern experiments \cite{Fu_2009}. Further, the analysis above has been
performed at zero temperature. At finite temperature $T$ there will appear a
noninteracting (or ideal) phonon gas. The average energy of a phonon in this
gas is of order $k_\text{B}T$. As soon as the energy of the phonons in the
CS exceeds $k_\text{B}T$, the sound distribution will not be
affected by thermal phonons. Since $k\thicksim 2p$ (see the previous section)
and for $b=200.0$ we have $p\approx 1.28\cdot
10^{-25}$ kg$\cdot$m/s, we get from the condition
$k_\text{B}T_0=ck$ that for temperatures $T<T_0\approx 53\text{K}$ the CS will
not be affected by thermal phonons. At higher temperatures the sharp Cherenkov
peaks shown in Fig. \ref{figure_5} should start to wash out. Finally, our
assumption of the isotropic Debye model has a little effect on the results
presented above. First, for Bi$_2$Te$_3$ the Debye wave vector is
$k_\text{D}\approx 1.3\AA^{-1}$, {\it i.e.}, it is much above the magnitudes
of the phonon wave vectors in the CS. Second, the phonon anisotropy,
$c_l/c_t\approx 1.8$, is much weaker than the CS anisotropy (several orders of
magnitude). Therefore, the contribution of the phonon anisotropy to the total
anisotropy of the CS will be negligible.

The estimate above clearly demonstrates that the CS on a surface of a 3D TI
and its unique features explored here may really be accessed and utilized in
modern experiments and future electronic devices based on 3D TI. Another
aspect making the CS fundamentally important is dissipation unavoidable in
realistic devices coupled to external environments. The CS is a ubiquitous
dissipative mechanism which, as follows from our estimate above, may determine
the efficiency of electronic devices based on helical particles.

\section{Acknowledgments}
Support from the DFG under the program SFB 689 is acknowledged.

\end{document}